\documentclass{cupconf}
\usepackage{psfig}

\title[Low frequency radio astronomy from the moon]{Low frequency
radio astronomy from the moon: cosmic reionization and more}

\author[Carilli, Hewitt, Loeb]
{C.L.Carilli$^1$, J.N. Hewitt$^2$, A.Loeb$^3$}

\affiliation{$^1$National Radio Astronomy Observatory, Socorro, 
NM, USA\\[\affilskip]
$^2$Kavli Institute for Astrophysics and Space
Research, Massachusetts Institute of Technology, 
Cambridge, MA\\[\affilskip]
$^3$Department of Astronomy, Harvard University, Cambridge, MA }

\pubyear{}
\volume{}
\pagerange{}
\date{\today}
\begin{document}

\maketitle

\begin{abstract}

We discuss low frequency radio astronomy from the moon, predominantly
in the context of studying the neutral intergalactic medium during
cosmic reionization using the HI 21cm line of neutral hydrogen.  The
epoch of reionization is the next frontier in observational cosmology,
and HI 21cm studies are recognized as the most direct probe of this
key epoch in cosmic structure formation.  Current constraints on
reionization indicate that the redshifted HI 21cm signals will likely
be in the range of 100 MHz to 180 MHz, with the pre-reionization
signal going to as low as 10 MHz. The primary observational challenges
to these studies are: 
\begin {itemize}
\item ionospheric phase fluctuations
\item terrestrial radio frequency interference
\item Galactic and extragalactic foreground radiation. 
\end{itemize}
\noindent Going to the far side of the moon
removes the first two of these challenges. Moreover, a low frequency
telescope will be relatively easy to deploy and maintain on the moon,
at least compared to other, higher frequency telescopes.  We discuss
the potential 21cm signals from reionization, and beyond, and the
telescope specifications needed to measure these signals. We then
describe ground-based projects currently underway to study the HI 21cm
signal from cosmic reionization. 
We include a brief discussion of other very low frequency
science enabled by being outside the Earth's ionosphere.

The near-term ground-based projects will act as path-finders for a
potential future low frequency radio telescope on the moon, both in
terms of initial scientific results on the HI 21cm signal from cosmic
reionization, and in terms of the telescope design and observing
techniques required to meet the stringent sensitivity and dynamic
range requirements. If it is found that the terrestrial interference
environment, or ionospheric phase fluctuations, preclude ground-based
studies of reionization, then it becomes imperative to locate future
telescopes on the far side of the moon. Besides pursuing these
path-finder reionization telescopes, we recommend a number of
near-term studies that could help pave the way for low frequency
astronomy on the moon.

\end{abstract}

\firstsection % if your document starts with a section,

\section{Introduction}

There is long standing interest in building a low frequency radio
telescope on the far side of the moon (Gorgolewski 1965; Burke 1985;
Kuiper et al. 1990; Burns \& Asbell 1991; Woan et al. 1997). The
reasons are clear (section 2): no ionosphere, and sheilding from
terrestrial radio frequency interference (RFI). Two factors have acted
to rekindle interest in low frequency radio astronomy from
moon. First, scientifically, efforts to study cosmic reionization
through the redshifted HI 21cm line have spurred numerous ground-based
low frequency projects (section 3).  And second is the new NASA
initiative to return Man to the moon, and beyond.

In this paper we discuss the advantages of building a radio telescope
on the far side of the moon. We present the primary scientific driver
in low frequency astronomy today, namely HI 21cm studies of cosmic
reionization, and discuss some of the near-term ground-based
telescopes being designed for these studies.  We also present other
very low frequency science programs that are enabled by going to the
moon. We close with a few recommendations for near-term studies that
might prove useful for planning a low frequency lunar telescope.

\section{Why the moon for low frequency radio astronomy?}

We start with the main advantages for considering the far side of the
moon for a low frequency radio telescope.  Low frequencies, in the
context of this paper, means frequencies, $\nu < 200$MHz ($\lambda
>1.5$m). In this regime it becomes cheaper, and possibly more
effective, to build  arrays of dipoles, which are
electronically steered through phasing of the dipole elements, as
opposed to steerable parabolic reflectors.

\begin{itemize} 

\item {\sl Ionospheric opacity:} The plasma frequency of the Earth's
ionosphere varies between roughly 10MHz and 20MHz, making the
ionosphere optically thick at lower frequencies. For the reionization
experiments discussed in Section 3, the relevant frequency range is $>
30$ MHz, hence ionospheric opacity is not an issue. We discuss in
Section 4 some very low frequency science, in the range 0.1 MHz and
10MHz, that are enabled by going to the moon. 

\item {\sl Ionospheric phase errors:} The fluctuating ionosphere also cause
variations in electronic path-length, which increase as
$\nu^{-2}$. Hence, fluctuations in the electron content of the
ionosphere will preclude low frequency imaging with synthesis arrays
unless a correction can be made for electronic phase variations due to
the varying ionosphere.  Figure 1 shows an example of ionospheric
phase errors on source positions using VLA data at 74 MHz (Cotton et
al. 2004; Lane et al. 2004).  Positions of five sources are shown for
a series of snap shot images over 10 hours.  A number of interesting
phenomena can be seen. First, the sources are slowly moving in
position over time, by $\pm 50"$ over timescales of hours. These
position shifts reflect the changing electronic path-length due to the
fluctuating ionosphere (ie.  tilts in the incoming wavefront due to
propagation delay).  Second, the individual sources move roughly
independently. This is a demonstration of the 'isoplanatic patch'
problem, ie.  the excess electrical path-length is different in
different directions.  At 74 MHz, the typical coherent patch size is
about 3$^o$ to 4$^o$. Celestial calibrators further than this distance
from a target source no longer give a valid solution for the combined
instrumental and propagation delay term required to image the target
source.  And third, at the end of the observation there occurs an
ionospheric storm, or traveling ionospheric disturbance, which
effectively precludes coherent imaging during the event.

New wide field self-calibration techniques, involving multiple phase
solutions over the field, or a `rubber screen' phase model (Cotton et
al. 2004; Hopkins et al. 2003), are being developed that should allow
for self-calibration over wide fields. Again, the moon presents a
clear advantage in this regard, being beyond the Earth's ionosphere.

\item {\sl Terrestrial radio frequency interference (RFI)}: Another problem
facing low frequency radio astronomy is terrestrial (man-made)
interference. Frequencies $< 200$MHZ are not protected bands, and
commercial allocations include everything from broadcast radio and
television, to fixed and mobile communications. At the lowest
frequencies ($< 1$MHz) the Earth's auroral emission dominates.

Many groups are pursuing methods for RFI mitigation and excision (see
Ellingson 2004). These include: (i) using a reference horn, or one
beam of a phased array, for constant monitoring of known, strong, RFI
signals, (ii) conversely, arranging interferometric phases to produce
a null at the position of the RFI source, and (iii) real-time RFI
excision using advanced filtering techniques in time and frequency, of
digitized signals both pre- and post-correlation. The latter requires
very high dynamic range (many bit sampling), and very high frequency
and time resolution.

In the end, the most effective means of reducing interference is to go
to the remotest sites.  A number of the near-term low frequency
path-finder telescopes (see Section 4), have selected sites in remote
regions of Western Australia and China, because of known low RFI
environments.

Clearly the best location to avoid terrestrial interference is the
far-side of the moon. Figure 2 shows the effect of the lunar radio
shadow on the RAE2 lunar orbiter (Alexander et al. 1975). This orbiter
had a low frequency ($< 10$MHz) radio receiver. The figure shows
complete blockage of the Earth's auroral emission during
immersion. Note that this interference blockage is the key argument
for the far side of the moon, as opposed to a free-flying space radio
telescope.

\item {\sl Ease of deployment and maintenance:} While not a scientific
rationale, it should be pointed out that a low frequency telescope may
be the easiest astronomical facility to deploy and maintain on the
moon. The antennas and electronics are high tolerance, with
wavelengths $> 1.5$m and system noise characteristics dominated by
the Galactic foreground radiation. Deployment could be automated,
using either javelin deployment (EADS/ASTRON), rollout of thin
polyimide films with metalic deposits (ROLSS; Lazio et al. 2006),
inflatable dipoles (LUDAR; Corbin et al. 2005), or deployment by
rovers. Likewise, being a phased array, low frequency telescopes are
electronically steered, and hence have no moving parts. Lastly,
there is no potential difficulty with lunar dust affecting the
optics. 

\end{itemize}

\begin{figure}
\psfig{figure=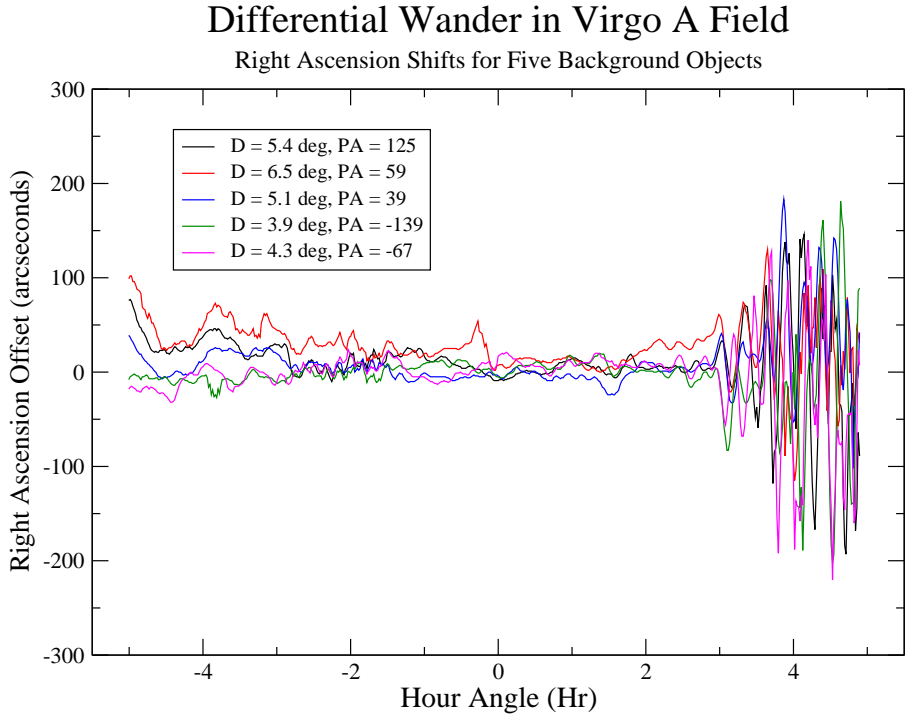,width=5.5in,angle=-90}
\caption{The positions of five sources in the Virgo A field at 74 MHz
observed with the VLA over 10 hours (Cotton et al. 2004; Lane et
al. 2004). The source positions vary with time due to fluctuations in
the electronic path-length through the ionosphere.  } \label{}
\end{figure}

\begin{figure}
\psfig{figure=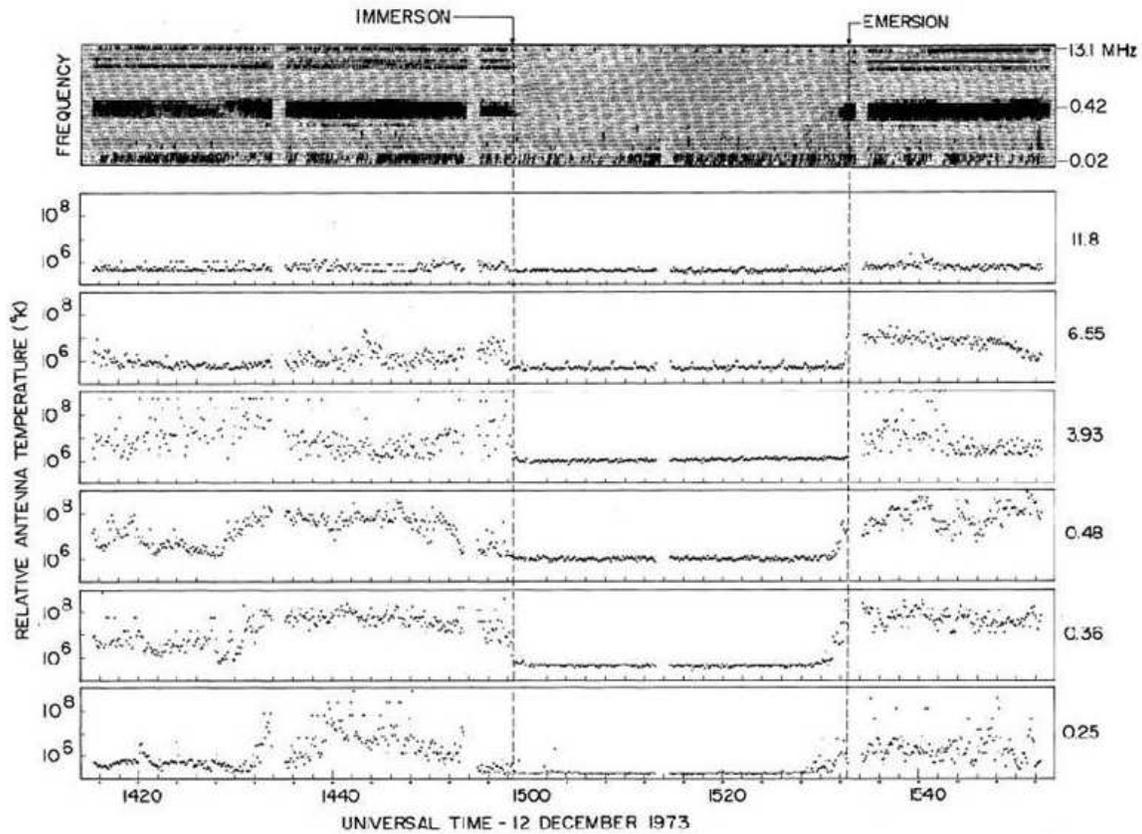,width=6in}
\caption{Radio power received by the lunar orbiting radio
astronomy explorer satellite in 1973 (RAE2; Alexander et al. 1975).
The power is dominated by the Earth's auroral emission except 
during immersion, when the Earth is totally blocked by the moon.
} \label{}
\end{figure}

\section{Cosmic Reionization}

\subsection{The HI 21cm signal} 

Cosmic reionization corresponds to the transition from a fully neutral
intergalactic medium (IGM) to an (almost) fully ionized IGM caused by
the UV radiation from the first stars and  blackholes.
Reionization is a key benchmark in cosmic structure formation,
indicating the formation of the first luminous objects.  Reionization,
and the preceding 'dark ages', represent the last of the major phases
of cosmic evolution to explore. Recent observations of the
Gunn-Peterson effect, ie.  Ly-$\alpha$ absorption by the neutral IGM,
toward the most distant quasars ($z \sim 6$), and the large scale
polarization of the CMB, have set the first constraints on the epoch
of reionization.  These data, coupled with the study of high-$z$
galaxy populations and other observations, suggest that reionization
was a complex process, with significant variance in both space and
time, starting perhaps as high as $z \sim 14$, with the last vestiges
of the the neutral IGM being etched-away by $z \sim 6$ (Fan et
al. 2006;, Ciardi \& Ferrara 2005; Loeb 2006).

The most direct and incisive means of studying cosmic reionization is
through the 21cm line of neutral Hydrogen (Furlanetto et al. 2006).
The study of HI 21cm emission from cosmic reionization entails the
study of large scale structure (LSS). During this epoch the entire IGM
may be neutral, and the LSS in question is not simply mass clustering,
but involves a combination of structure in cosmic density, neutral
fraction, and HI excitation temperature (Loeb 2006).  Hence, HI 21cm
observations are potentially the `richest of all cosmological data
sets' (Loeb \& Zaldarriaga 2004). Information about the physics of
cosmology (involving the initial perturbations from inflation and the
matter content of the universe) can be separated from the astronomical
aspects (involving galaxy formation) through the line-of-sight
anisotropy imprinted by peculiar velocities on the power-spectrum of
21cm brightness fluctuations (Barkana \& Loeb 2005a).  We briefly
describe a few of the potential HI 21cm signatures of cosmic
reionization.

\begin{figure}
\psfig{file=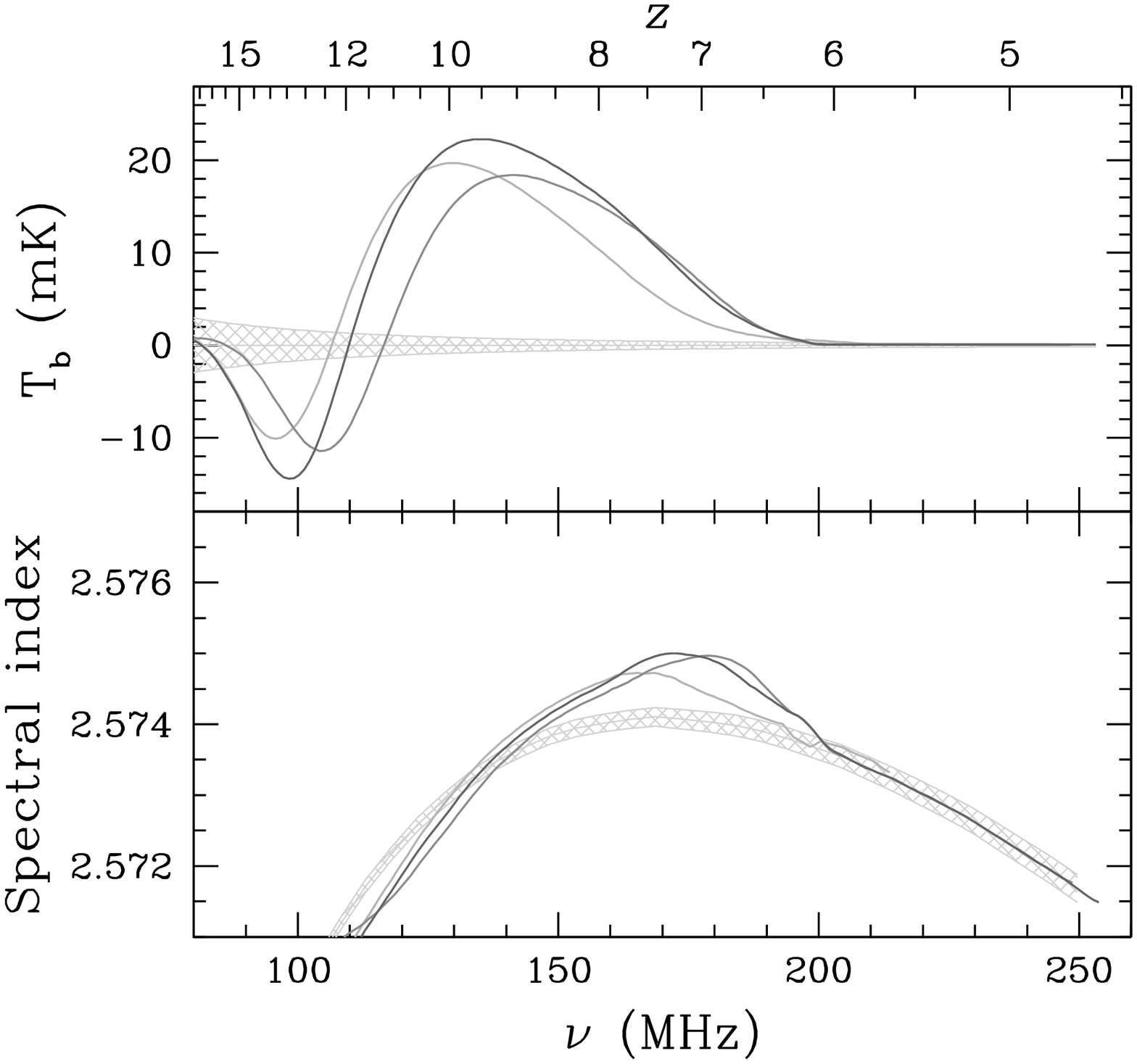,width=2.0in}
\vskip -1.8in
\hspace*{2.1in}
\psfig{file=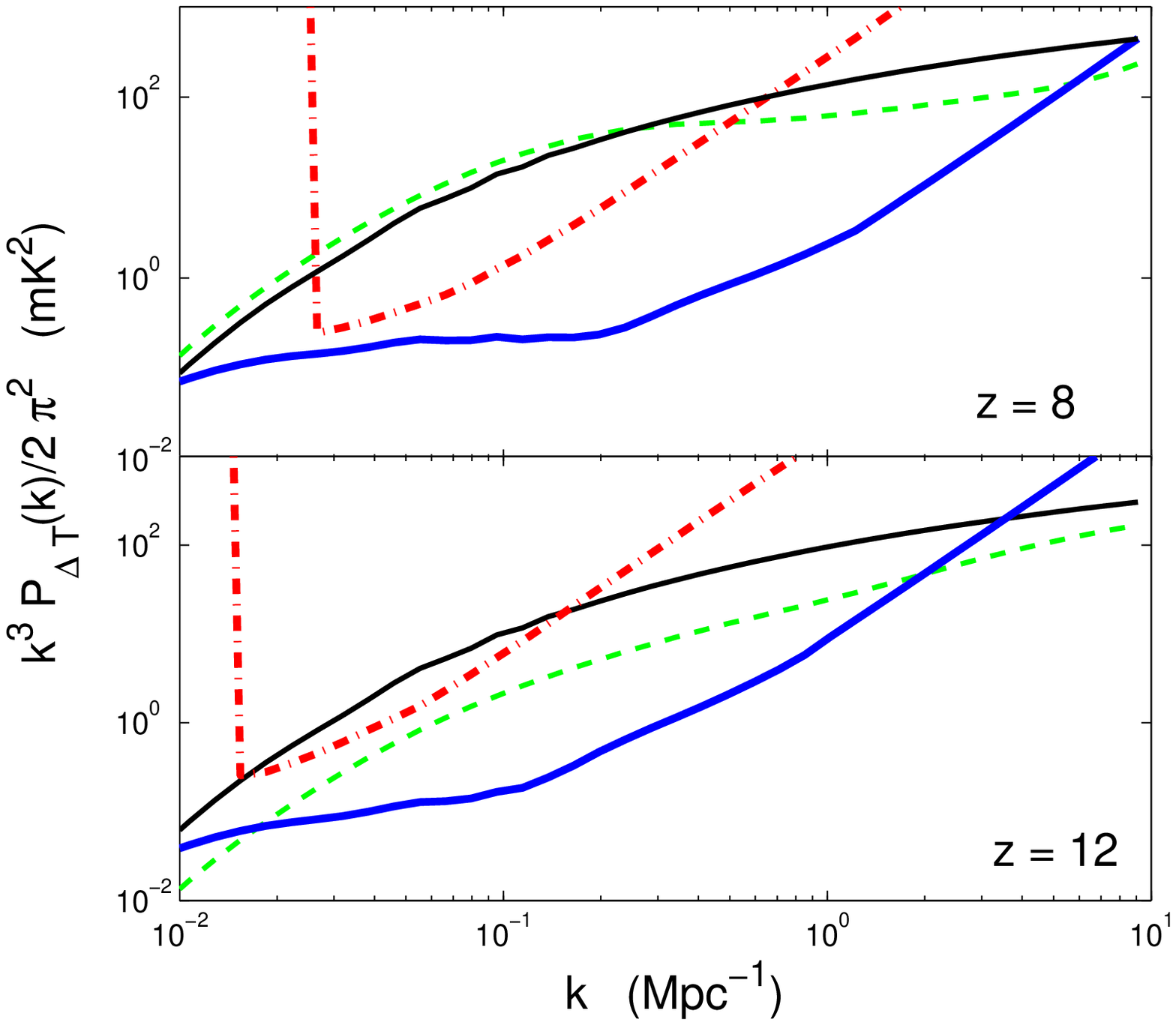,width=2.0in}
\vskip -1.82in
\hspace*{4.2in}
\psfig{file=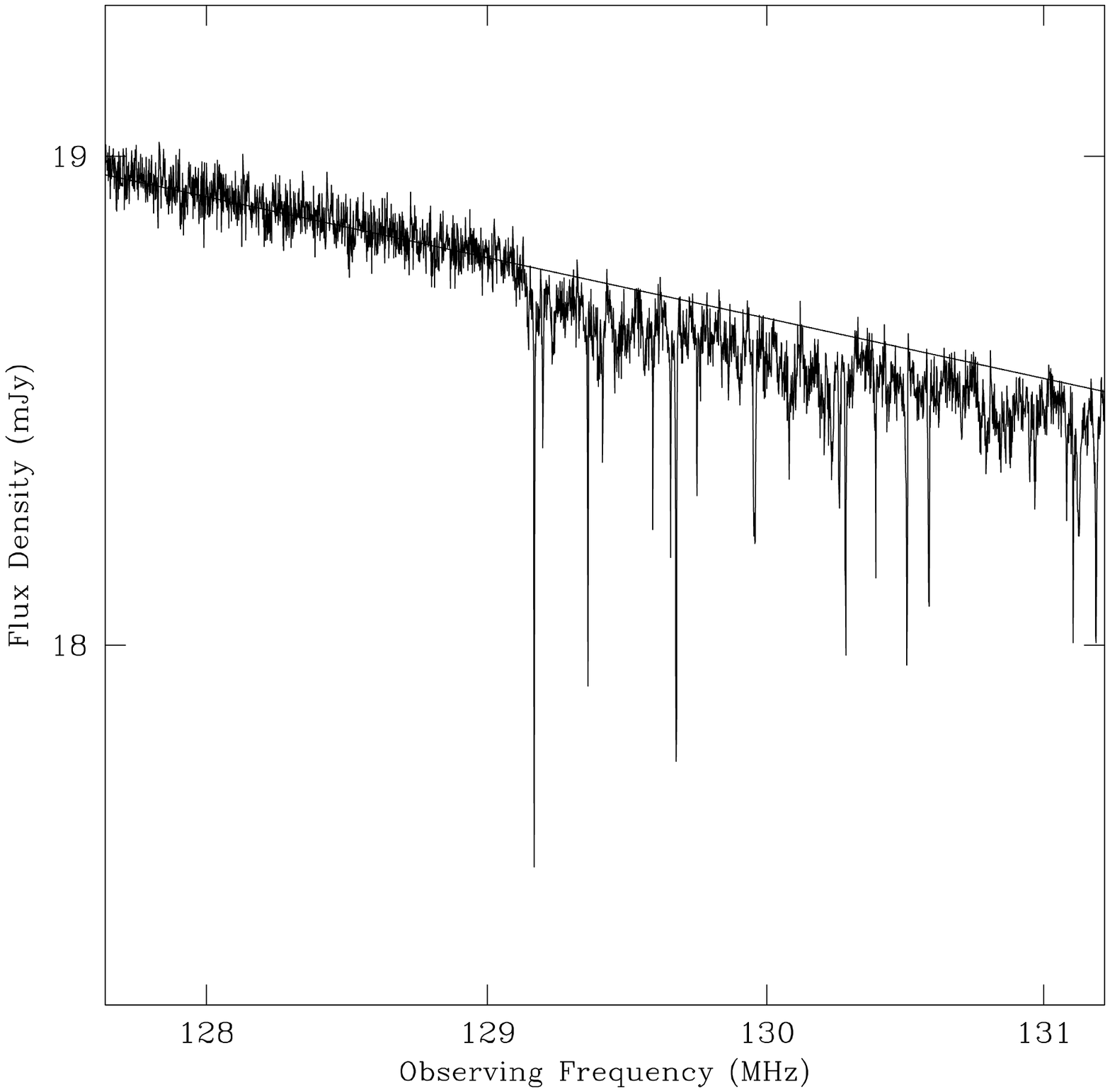,width=1.8in}
\hskip 0.2in
\caption{{\sl Left}: Global (all sky) HI signal from reionization
(Gnedin \& Shaver 2003). 
The shaded region shows the expected thermal
noise in a carefully controlled experiment.
{\sl Center:} Predicted HI 21cm brightness temperature power spectrum
 (in log bins) at redshifts 8 and 12 
(Mcquinn et al. 2006).   The thin
black line shows the signal when density fluctuations dominate.  The
dashed green line shows the predicted signal for ionization fraction, 
$\bar{x}_i = 0.2$
at $z=12$, and $\bar{x}_i = 0.6$ at $z=8$,
in the Furlanetto et al. (2004) semi-analytic model. 
The thick blue line
shows the Square Kilometer Array (SKA; Carilli \& Rawlings 2004) 
sensitivity in 1000hrs. The thick red dot-dash show the
sensitivity of the path-finder experiment LOFAR. The cutoff at low k is
set by the primary beam.
{\sl Right}:  The simulated SKA spectrum of a radio continuum source
at $z = 10$ (Carilli et al. 2002). 
The straight line is the intrinsic
power law (synchrotron) spectrum of the source. The noise curve
shows the effect of the 21cm line in the neutral IGM, including
noise expected for the SKA in a 100 hour integration.}
\end{figure}

\begin{itemize}
 
\item {\sl Global signal:} The left panel in Figure 3 shows the latest
predictions of the global (all sky) increase in the background
temperature due to the HI 21cm line from the neutral IGM (Gnedin \&
Shaver 2003).  The predicted HI signal peaks at roughly 20 mK above
the foreground at $z \sim 10$. At higher redshift, prior to IGM
warming, the kinetic temperature of the IGM will be colder than the
CMB temperature. In this case, Ly$\alpha$ emission from the first
luminous objects can couple the gas kinetic temperature to the HI spin
temperature through the resonant scattering of Ly$\alpha$ photons
(Wouthuysen 1952; Field 1959). In this case, the HI will be seen in
absorption against the CMB.  Since this is an all sky signal, the
sensitivity of the experiment is independent of telescope collecting
area, and the experiment can be done using small area telescopes at
low frequency, with very well controlled frequency response (Rogers \&
Bowman, in prep; Subrahmanyan, in prep).  Note that the line signal is
only $\sim 10^{-4}$ that of the mean foreground continuum emission at
$\sim 150$ MHz (see Section 3.2).

\item {\sl Power spectra:} The middle panel in Figure 3 shows the predicted
power spectrum of spatial fluctuations in the sky brightness
temperature due to the HI 21cm line (Mcquinn et al. 2006).  For power
spectral analyses the sensitivity is greatly enhanced relative to
direct imaging due to the fact that the universe is isotropic, and
hence one can average the measurements in annuli in the Fourier (uv)
domain, ie. the statistics of fluctuations along an annulus in the
uv-plane are equivalent.  Moreover, unlike the CMB, HI line studies
provide spatial and redshift information, and hence the power spectral
analysis can be performed in three dimensions. The rms fluctuations at
$z = 10$ peak at about 10 mK rms on scales $\ell \sim 5000$ ($\sim
2'$).

\item {\sl Absorption toward discrete radio sources:} A interesting
alternative to emission studies is the possibility of studying smaller
scale structure in the neutral IGM by looking for HI 21cm absorption
toward the first radio-loud objects (AGN, star forming galaxies, GRBs)
(Carilli et al. 2002). The right panel of Figure 3 shows the predicted
HI 21cm absorption signal toward a high redshift radio source due to
the `cosmic web' prior to reionization, based on numerical
simulations.  For a source at $z = 10$, these simulations predict an
average optical depth due to 21cm absorption of about 1$\%$,
corresponding to the `radio Gunn-Peterson effect', and about five
narrow (few km/s) absorption lines per MHz with optical depths of a
few to 10$\%$. These latter lines are equivalent to the Ly $\alpha$
forest seen after reionization. Furlanetto \& Loeb (2002) predict a
similar HI 21cm absorption line density due to gas in
minihalos\footnote{Cosmic mini-halos are the first objects with masses
large enough to overcome the cosmic Jeans mass and collapse,
ie. masses between $10^6$ M$_\odot$ and $10^7$ M$_\odot$.}, as that
expected for the 21cm forest. This absorption experiment may be the
easiest of the HI 21cm probes of reionization to perform, since it
entails absorption against a high brightness temperature object, and
hence is less affected by 3D dynamic range issues, and it can use long
baselines, which are less susceptable to interference. However, the
experiment is predicated on the existence of radio sources at these
during reionization.  This question has been considered in detail by
Carilli et al. (2002), Haiman et al. (2004), and Jarvis \& Rawlings
(2005).  They show that current models of radio-loud AGN evolution
predict between 0.05 and 1 radio sources per square degree at $z > 6$
with $\rm S_{150MHz} \ge 6$ mJy, adequate for reionization HI 21cm
absorption studies with the Square Kilometer Array (SKA).

\item {\sl Tomography:} Figure 4 shows the expected evolution of the HI 21cm
signal during reionization based on numerical simulations (Zaldarriaga
et al. 2004).  In this simulation, the HII regions caused by galaxy
formation are seen in the redshift range $z \sim 8$ to 10, reaching
scales up to 2$'$ (frequency widths $\sim 0.3$ MHz $\sim 0.5$ Mpc
physical size). These regions appear as 'holes in the sky', with
(negative) brightness temperatures up to 20 mK. This corresponds to
5$\mu$Jy beam$^{-1}$ in a 2$'$ beam at 140 MHz. Only a full square
kilometer of collecting will be able to perform the 3D tomographic
imaging of the typical structures during reionization (Section 3.3).

\item {\sl Cosmic Stromgren spheres:} While direct detection of the typical
structure of HI and HII regions may be out of reach of the path-finder
telescopes (Section 3.2), there is a chance that even these first
generation telescopes will be able to detect the rare, very large
scale HII regions associated with luminous quasars near the end of
reionization.  The expected signal is $\sim 20$mK $\times ~ x_{HI}$ on
scales $\sim 10'$ to 15$'$, with line widths $\sim 1$ to 2 MHz (Wyithe
et al. 2005), where $x_{HI}$ is the IGM neutral fraction, by volume.
This corresponds to 0.5 $\times ~ x_{HI}$ mJy beam$^{-1}$, for a 15$'$
beam at $z \sim 6$ to 7.

\end{itemize}

\begin{figure}
\centerline{\psfig{file=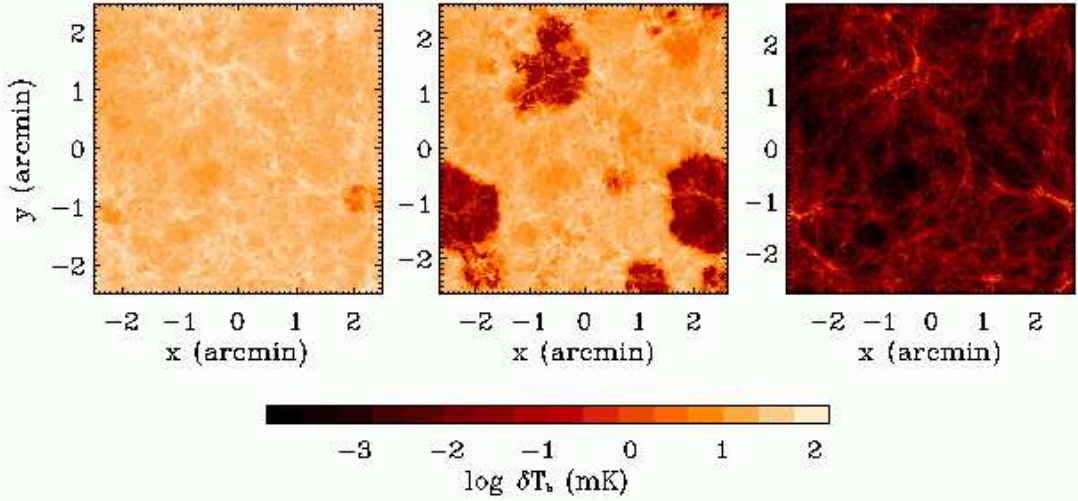,width=6in}}
\caption{The simulated HI 21cm brightness temperature distribution
during reionization at $z = 12$, 9, 7 (left to right; 
Zaldarriaga et al. 2004).}
\label{HIsim}
\end{figure}

\subsection{Sensitivities, foregrounds, and telescope requirements}

A number of groups have calculated the expected HI 21cm signals from
cosmic reionization, including large bubbles associated with bright
quasars, and clustering of star forming galaxies (Zaldarriaga et al.
2004; Wyithe et al. 2005; Mellema et al. 2006; Madau et
al. 1997).  The larger structures will vary from a few to 15 arcmin in
size, with a depth of $20\times ~ x_{HI}$ mK. For demonstrative
purposes, we assume a target signal for detection of 10mK and 10$'$ in
size at $z = 8$. This implies an observing frequency of $\nu =
158$MHz, or observing wavelength, $\lambda = 1.9$m. For reference, at
$z \sim 8$, 10$'$ = 3.2 Mpc physical size, or in co-moving coordinates
10$'$ = 3.2/(1+z) = 0.36 Mpc co-moving, and in terms of the Hubble
expansion, 3.2 Mpc (physical) = 1.6 MHz. The angular size of 10$'$
corresponds to a baseline length of 650m at 158MHz.

The relationship between brightness temperature and flux density
is given by: 

$$\rm T_B = 1360 {{S_\nu}\over{\theta^2}} \lambda^2 ~ K$$

\noindent where S$_\nu$ is the flux density in Jy, $\theta$ is the
angular size, in arcseconds, and $\lambda$ is the observing wavelength
in centimeters.  For reference,  a 10mK signal at $\lambda = 1.9$m 
and $\theta = 10'$ corresponds to 73$\mu$Jy.

{\sl Sensitivity:} The sensitivity of a radio telescope is given by
the radiometry equation:

$$\rm rms = 3000 {{T_{sys}}\over{\epsilon N A}} (\Delta \nu
~t)^{-1/2} ~ Jy $$

\noindent where $\epsilon$ is the antenna efficiency, N is the number
of antennas, A is the physical collecting area of each antenna, in
m$^2$, $\Delta \nu$ is the bandwidth (Hz), and t is the integration
time (seconds). The system temperature, $\rm T_{sys}$, is the sum of
the receiver temperature, which we assume is of order 100 K, and the
contribution from the sky foreground. The sky is about 90\% diffuse
Galactic emission, and 10\% extragalactic radio sources.  In the
coldest regions, the sky contributes:

$$\rm T_{sky} \sim 100 ({{\nu}\over{200MHz}})^{-2.8}$$

\noindent or about 200K at 158MHz. In general, at low frequencies
the sky brightness dominates the system temperature.  Note that the
expected signal is roughly $2\times 10^4$ times weaker than the
diffuse foreground brightness temperature, even in the coldest regions
of the sky.

What collecting area is needed to detect the 73$\mu$Jy signal at
4$\sigma$ at 158MHz in 1000 hrs, assuming a bandwidth of 1.6MHz and an
efficiency of 60\%?  The required area = $3.4\times 10^4$ m$^2$. This
corresponds to roughly 9500 dipoles, assuming each dipole collecting
area is of order $\lambda^2$.

The path-finder arrays will likely not have the sensitivity to perform
true 3D 'tomographic imaging' of the HI 21cm signal during
reionization. These arrays are being designed to study the statistical
signal, ie. the power spectrum of brightness temperature fluctuations
due to the structure of the neutral IGM, in a manner similar to the
COBE statistical detection of CMB brightness temperature fluctuations
(see section 3.1). The sensitivity for these near-term arrays to study
the power spectrum of the neutral IGM during reionization is shown in
Figure 3.

{\sl Dynamic range:} Besides raw sensitivity, a significant challenge
to wide field, low frequency interferometric imaging is dynamic range,
set by residual calibration errors and incomplete Fourier spacing
coverage. Besides the diffuse Galactic foreground, every field will
have bright extragalactic and Galactic continuum sources.

As a rough estimate for the number of sources expected in a given
field, we adopt a field size of $20^o \times 20^o$, corresponding
to the field of view of a receptor 'tile' of $3 \times 3$ dipoles. 
Using the 3C catalog, the bright source counts at 1.9m follows roughly:

$$\rm N(> S) = 0.29 ~ S^{-1.36} ~ deg^{-2}$$

\noindent where N is the number of sources per deg$^{-2}$ brighter
than S in Jy. Hence, in a typical 400 deg$^2$ region, we expect
one source brighter than 34 Jy at 158 MHz. 

The dynamic range requirement $\equiv$ (Peak in field)/(rms required).
The rms required for a 4$\sigma$ detection = 73/4 = 18 $\mu$Jy.
Hence, the DNR = 34 Jy/18 $\mu$Jy = $1.9\times 10^6$.

Perley (1999; equ. 13-8) derives the dynamic range limit for a
synthesis array assuming 'random' antenna based phase errors, $\Delta
\phi$ (in radians):

$$ \rm DNR \sim {{N}\over{2^{1/2} \Delta \phi}} $$

\noindent where, again, N is the number of elements in the array. 
Assuming 9500 dipoles grouped in 3x3 tiles implies N = 1055 elements.
The requirement then on the phase calibration is: 
{\sl $\Delta \phi < 0.023^o$}. 

There have been numerous studies on how to separate the HI 21cm signal
from the foreground continuum emission at the required levels.  All
these techniques rely on the spectral nature of the HI signal. The
foreground emission is syncrotron radiation, and hence shows at most
gradual changes with frequency on scales of tens to hundreds of
MHz. The HI signal is a resonant spectral line, and will have
structure on scales of kHz to a few MHz.  A number of complimentary
approaches have been presented for foreground removal (Morales,
Bowman, \& Hewitt 2005).  Gnedin \& Shaver (2003) and Wang et al
(2005) consider fitting smooth spectral models (power-laws or low
order polynomials in log space) to the observed visibilities or
images. Morales \& Hewitt (2003) and Morales (2004) present a 3D
Fourier analysis of the measured visibilities, where the third
dimension is frequency.  The different symmetries in this 3D space for
the signal arising from the noise-like HI emission, versus the smooth
(in frequency) foreground emission, can be a powerful means of
differentiating between foreground emission and the reioniation HI
line signal.  Santos et al. (2005), Bharadwaj \& Ali (2005), and
Zaldarriaga et al. (2004) perform a similar analysis, only in the
complementary Fourier space, meaning cross correlation of spectral
channels. They show that the 21cm signal will effectively decorrelate
for channel separations $> 1$ MHz, while the foregrounds do not.  The
overall conclusion of these methods is that spectral decomposition
should be adequate to separate synchrotron foregrounds from the HI
21cm signal from reionization at the mK level, as long as residual, at
least in the absence of residual significant, frequency dependent
calibration errors.

\subsection{Telescopes}

Table 1 summarizes the current experiments under construction to study
the HI 21cm signal from cosmic reionization.  These experiments vary
from single dipole antennas to study the all-sky signal, to 10,000
dipole arrays to perform the power spectral analysis, and potentially
to image the largest structures during reionization (eg. the quasar
Stromgren spheres).

Most of the experiments have a few to 10\% of the collecting area of
the SKA, and there are many common features.  First, they all rely on
some form of a wide-band dipole or spiral antenna, eg.  log-periodic
yagis, sleeve dipoles, or bow-ties, with steering of the array
response through electronic phasing of the elements.  Second, the
front-end electronics are relatively simple (amplifier/balun), since
the system performance is dominated by the sky brightness temperature.
Third, most rely on a grouping of dipoles into 'tiles' or 'stations',
to decrease the field-of-view, and to decrease the data rate into the
correlator to a managable level.  And forth, the large number of array
elements, and the need for wide-field, high dynamic range imaging over
an octave, or more, of bandwidth, demand major computing resources,
both for basic cross-correlation, and subsequent imaging and
analysis. For example, assuming an array of 1000 tiles, a 100 MHz
bandwidth, and 8 bit sampling, the total data rate coming into the
correlator is 1.6 Tbit s$^{-1}$. The LOFAR array is working with IBM
to apply the 27.4 Tflop Blue Gene supercomputing technology to
interferometric imaging (Falcke 2006).

Figure 5 shows one array under construction.  The 21cm Chinese Meter
Array (21CMA) involves of order 10$^4$ Yagis in western
China\footnote{http://cosmo.bao.ac.cn/project.html}.  First results
from these path-finder telescopes are expected within the next few
years.  Experience from these observations in dealing with the
interference, ionosphere, and wide-field imaging/dynamic range
problems will provide critical information for future experiments,
such as the Square Kilometer Array, or a low frequency radio telescope
on the moon.

\begin{figure}
\centerline{\psfig{file=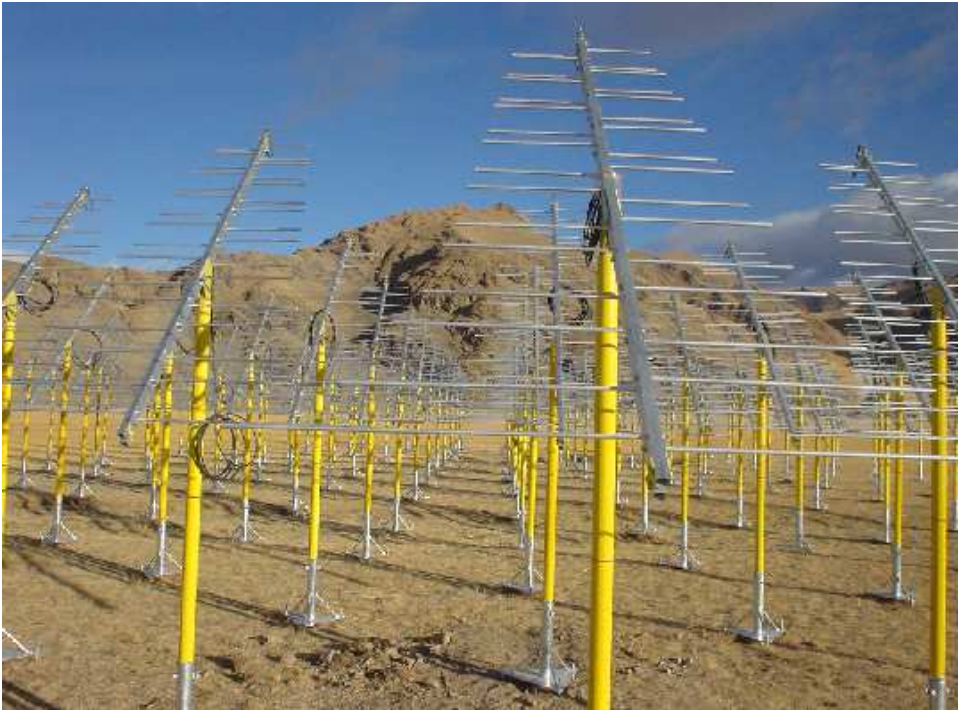,width=5in}}
\caption{The 21cm Chinese Meter Array
(21CMA) in China (http://cosmo.bao.ac.cn/project.html). 
}
\end{figure}

\begin{table}
  \begin{center}
  \begin{tabular}{lcccccc}
experiment & site & type & $\nu$ range & Area & date & goal \\
~ & ~ & ~ & MHz & m$^2$ & ~ & ~ \\
\hline
\hline
Mark I$^a$ & Australia & spiral & 100-200 & few & 2007 & All Sky \\
EDGES$^b$ & Australia & four-point & 100-200 & few & 2007 & All Sky \\
GMRT$^c$ & India & parabola array & 150-165 & 4e4 & 2007 & CSS$^d$ \\
PAPER$^e$ & Australia & dipole array & 110-190 & 1e3 & 2007 & PS/CSS/Abs \\
21CMA$^f$ & China & dipole array & 70-200 & 1e4 & 2007 & PS \\
MWAd$^g$  & Australia & aperture array & 80-300 & 1e4 & 2008 & PS/CSS/Abs \\
LOFAR$^h$ & Netherlands & aperture array & 115-240 & 1e5 & 2008 & PS/CSS/Abs \\
SKA$^i$ & ? & aperture array & 100-200 & 1e6 & 2015 & Imaging \\
\hline
  \end{tabular}
  \caption{HI 21cm Cosmic Reionization Experiments}
  \label{tab:tab}
  \end{center}
$^a$http://www.atnf.csiro.au/news/newsletter/jun05/Cosmological\_re-ionization.htm \\
$^b$http://www.haystack.mit.edu/ast/arrays/Edges/index.html \\
$^c$http://gmrt.ncra.tifr.res.in/ \\
$^d$CSS = cosmic Stromgren Spheres, PS = power spectrum, Abs = absorption \\
$^e$http://astro.berkeley.edu/$\sim$dbacker/EoR/ \\
$^f$http://cosmo.bao.ac.cn/project.html \\
$^g$http://www.haystack.mit.edu/ast/arrays/mwa/ \\
$^h$http://www.lofar.org/ \\
$^i$http://www.skatelescope.org/ \\
\end{table}

\subsection{Pre-reionization signal: the difficulty with probing
the 'dark ages'}

A number of studies have considered the pre-reionization HI 21cm
signal (Loeb \& Zaldarriaga 2004; Cen 2006; Barkana \& Loeb 2005b;
Shethi 2005).  The HI 21cm measurements can explore this physical
regime at $z \sim 50$ to 300, or redshifts $z > 30$. At these
frequencies, going to the moon becomes more imperative, due to the
rapidly increasing ionospheric opacity and phase effects.

In this redshift regime the HI generally follows linear density
fluctuations, and hence the experiments are as clean as CMB studies,
and $T_K < T_{CMB}$, so a relatively strong absorption signal might be
expected.  Also, Silk damping, or photon diffusion, erases structures
on scales $\ell > 2000$ in the CMB at recombination, corresponding to
comoving scales = 22 Mpc, leaving the 21cm studies as the only
current method capable of probing to very large $\ell$ in the
linear regime. The predicted rms brightness temperature
fluctuations are 1 to 10 mK on scales $\ell= 10^3$ to 10$^6$ (0.2$^o$
to 1$''$).  These observations could provide the best test of
non-Gaussianity of density fluctuations, and constrain the running
power law index of mass fluctutions to large $\ell$, providing
important tests of inflationary structure formation. Sethi (2005) also
suggests that a large global signal, up to -0.05 K, might be expected
for this redshift range.

The difficulty in this case is one of sensitivity. The sky temperature
is $> 10^4$K, and using the equations in section 3.2, it is easy to
show that, even if structures as bright at 10mK on scales of a few
arcminutes exist, it would require $\sim 10$ square kilometers of
collecting area to detect them. For the more typical small scale
structure being considered (ie. $\sim 10"$), the required collecting
area increases to $3.6\times 10^{10}$ m$^2$. The dynamic range
requirements also become extreme, $> 10^8$.  It should also be kept in
mind that the sky becomes highly scattered due to propagation through
interstellar and interplanetary plasma, with source sizes obeying:
$\rm \theta_{min} \sim 1 ({{\nu}\over{1 MHz}})^{-2}$deg. Hence the all
objects in the sky are smeared to $> 4"$ for frequencies $< 30$ MHz.

\section{Very low frequency science (1 MHz to 10MHz) from the moon}

A new astronomical window is opened up by going outside the Earth's
ionosphere, between 1 MHz and 10MHz. The lower limit of 0.1 MHz is
set by a combination of the heliospheric plasma frequency, and
Galactic free-free absorption.
We briefly discuss some interesting science opportunities generated by
opening up this window (see Burns this volume). This is clearly an
incompletely list, and a key point is that the most interesting
discoveries that come from opening a new astronomical window are
usually not predictable.

\begin{itemize} 

\item {\sl Coronal mass ejections (CME's) and space weather:} Solar magnetic
activity generates ionized mass ejections which can severely affect
satellites, and other electronic equiptment, eg. associated with a
lunar base.  These CME's can be studied at low radio frequencies
(Bastian 2006), both passively through various plasma radiation
processes, and through remote sensing (radar).  A low frequency radio
telescope could play an important role as part of a severe space
weather 'early warning system', allowing for appropriate action to be
taken for satellite and lunar base safety.

\item {\sl Planetary radio emission:} Bursts of low frequency emission from
Jupiter were discovered early-on by Franklin \& Burke (1956),
associated with the interaction between the Io-driven ion torus and
Jupiter's magnetic field.  Many other low frequency emission
mechanisms exist for planets, ranging from aurural emission associated
with the interaction of the solar wind and planetary magnetic fields,
to lightening in planetary atmostpheres (Woan et al. 1997).

\item {\sl Extrasolar planetary radio bursts:} Lazio et al. (2004) has
calculated the expected frequency and signal strength of Jupiter-burst
type radio emission from all extrasolar planets known at the time,
based on a coarse relationship between planet mass and radio power
(Figure 6).  A low frequency radio telescope has the potential to be
an effective planet-finding instrument through low frequency radio
bursts at the mJy level between 1 and 100 MHz.

\item {\sl Neutrino interactions with the lunar regolith:} High energy
neutrinos passing through the lunar regolith (from the far side), may
interact with the regolith, generating a shower of energetic particles
which emit a burst of beamed Cherenkov radiation. Attempts have been
made to use the moon as a neutrino detector through the resultant low
frequency radio emission (Hankins et al. 2000). A set of local
detectors on the moon could be used to study the emission {\sl in
situ} (Falcke 2006).

\item {\sl Synchrotron emission by intergalactic collisionless shocks:} The
shocks produced by converging flows in the intergalactic medium
accelerate a power-law tail of electrons to relativistic energies. The
synchrotron emission by these electrons in the post-shock magnetic
fields paints a cosmic web of radio emission and should be brightest
at low frequencies (Keshet et al. 2005). The emission is already seen
in the accretion shocks around X-ray clusters (Bagchi et al. 2006).

\end{itemize}

\begin{figure}
\centerline{\psfig{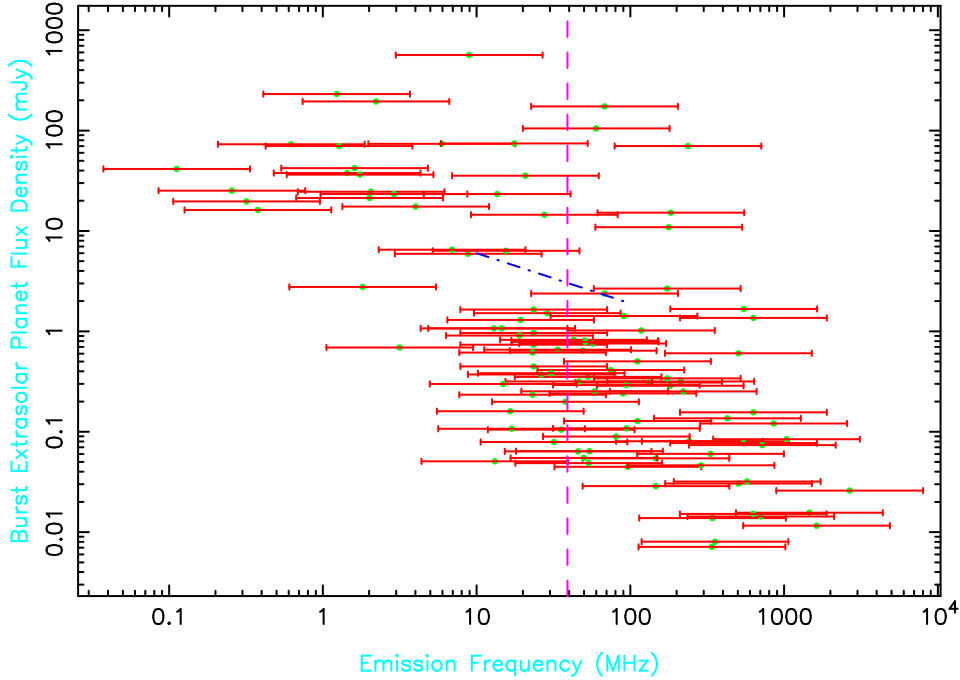}}
\caption{Predicted radio burst flux densities from all known 
extrasolar planets (Lazio et al. 2004), with 
approximate indication of the sensitivity (blue dash-dot line) of a
ground-based, long-wavelength telescope like the Long
Wavelength Array (Kassim et al. 2006), assuming a 15 minute integration.
}
\end{figure}

\section{Near-term lunar investigations}

There are a number of issues that could be investigated in preparation
for a lunar implementation of a low frequency radio telescope.

\begin{itemize} 

\item  {\it Lunar ionosphere:} Studies in the 1970's using the LUNA lunar
orbiters, suggested that the moon may have a significant ionosphere,
with a day side plasma frequency of about 1MHz and a night side value
of 0.2 MHz (Vyshlov 1974). The origin, or even existence, of this
ionosphere remains unclear (Bauer 1976), with possible sources being
radioactivity on the moon, capture of the solar wind, and ionization
by cosmic rays.  Even if real, the plasma frequency is still at least
an order of magnitude below the Earth's value, hence opacity is not an
issue. However, the issue of electronic path-length variations due to
the lunar ionosphere needs to be explored.  Experiments are needed to
test the existence of the lunar ionosphere, and if real, determine
if there are significant electronic path-length variations at
frequencies $< 200$ MHz.

\item {\it Computing requirements and power consumption:} We discussed in
section 3.3 that the data rates for a full-up array to study
reionization are extreme (1.6 Tb/s). The station data cannot be
transmitted directly to Earth, but need to correlated on the moon.
The power consumption is significant, eg. the Blue Gene correlator for
LOFAR draws 0.15MW of power. This raises the potential requirement of
low power supercomputing, and early design studies are required to
address this issue. Of course, for an initial small test array, the
data rate may be managable, eg. 8 elements with 200 MHz/8 bit sampling
implies a data rate of only 12.8 Gbit s$^{-1}$. On the timescales
being considered ($\sim 2020$), Moore's law for computing should also
provide significantly more capabilities than are currently available.

\item {\it RFI shielding:} Given the power requirements above, it has been
proposed to locate the array in a crater near the lunar pole
(Heidemann 2000; Woan et al 1997). Within the crater there can be a
region of permanent shadowing from the Earth, and yet on the crater's
rim there may be eternal sunlight. The question arises: how far around
the limb of the moon does the array have to be for adequate shielding
from terrestrial radio emission?

\item {\it Current arrays:} The current arrays will be key path-finders in
the study of element design, array design, foreground removal,
wide-field imaging with large fraction bandwidths, correlation of many
element interferometers, and related.  These arrays will also
demonstrate whether the requisit very high dynamic range images ($\sim
10^6$) can be generated in the presence of large ionospheric phase
distortions and strong terrestrial interference.  If not, then the far
side of the moon becomes imperative for future probes of this key
epoch of cosmic evolution.

\item {\it Form a lunar radio telescope working group}, to help coordinate
near-term studies related to a future lunar radio array, and provide
funding for preliminary studies related to a lunar radio telescopes.

\item {\it Enforce ITU agreement 22.22}, keeping the lunar far-side 
a radio quiet zone.

\end{itemize}

\section{Summary}

Cosmic reionization is the next frontier in observational cosmology.
The HI 21cm signal at low frequencies from the neutral IGM during
reionization is the most direct probe of the physical processes
involved in reionization. A number of low frequency, ground-based
path-finder arrays are being planned to make the first detection
of the neutral IGM during reionization. The far-side of the
moon presents the unique opportunities of being outside the Earth's 
ionosphere, and shielded from terrestrial interference. 

Beyond supporting the current ground-based efforts, we have proposed a
number of design, environment, or technical studies that could be
performed to pave the way to a lunar low frequency radio telescope.
Lunar development could be staged, while providing results of
scientific interest along the way. A simple experiment would be a
single low frequency dipole to study the all-sky reionization
signal. The next phases could involve arrays of similar aperture to
the current ground-based arrays, to study the statistical signal, and
other aspects, like absorption toward very high $z$ radio sources.
The final stage could entail a full square kilometer of collecting
area to perform 3D tomographic imaging of the neutral intergalactic
medium during cosmic reionization. We should point out that the
Europeans are well along in planning for the first automated low
frequency radio telescope on the moon, with a planned Ariane V launch
sometime in the coming
decade\footnote{http://www.astron.nl/p/lunar\_observatories.htm}.

\begin{acknowledgments} 
CC thanks the Max-Planck-Gesellschaft and the
Humboldt-Stiftung for partial support through the
Max-Planck-Forschungspreis.  AL was supported in part 
by Harvard University and FQXi grants. We thank the authors
of the papers referenced in the figure captions for permission
to reproduce their figure, and J. Lazio for comments.
\end{acknowledgments}

{}


\begin{thebibliography}{}

\bibitem[]{} Alexander, J. K., Kaiser, M. L., Novaco, J. C., Grena,
F. R., Weber, R. R. 1975, A\& A, 40, p. 365

\bibitem[]{} Bagchi, J., Durret, F.,
Neto, G.~B.~L., \& Paul, S.\ 2006, Science, 314, 791

\bibitem[]{} Barkana, R., Loeb, A. 2005a, ApJ, 624, L65-68

\bibitem[]{} Barkana, R. \& Loeb, A. 2005b, MNRAS, 363, L36

\bibitem[]{} Bastian, T. 2006, in {\sl From Clark Lake to the LWA},
eds. N. Kassim et al., San Francisco: ASP, p. 142

\bibitem[]{} Bauer, S. in {\sl Solar wind interaction with the
planets}, ed. N. Ness, NASA-SP-379, p. 47

\bibitem[Bharadwaj \& Ali 2005]{BA04}
Bharadwaj, S. \& Ali, Sk. 2005, MNRAS, 356, 1519-1428

\bibitem[]{} Burns, J.O. \& Asbell, J., 1991
in {\sl Radio astronomy from space}, NRAO:Greenbank,
ed. K. Weiler, p. 29

\bibitem[]{} Burke, B.F. 1985, in {\sl Lunar bases and space activity
of the 21st century}, LPI: Houston, ed. W. Mendell, p. 281

\bibitem[]{}
Field, G.B. 1959, ApJ, 129, 551-565

\bibitem[]{} Franklin, K. \& Burke, B. 1956, AJ, 61, 177

\bibitem[Carilli et al. (2002)]{CGO02}
Carilli, C., Gnedin, N., Owen, F. 2002, ApJ,
577, 22-30

\bibitem[]{} Carilli, C. \& Rawlings, S. 2004, NewAR, 48, 979

\bibitem[]{} Cen, R. 2006, ApJ, 648, 47

\bibitem[]{} Cotton, W., Condon, J., Perley, R. et al. 2004,
SPIE, 5489, 180-189

\bibitem[Cen 2003a]{cen2003a} Cen, R.\ 2003a, ApJ, 591, 12-37

\bibitem[Ciardi \& Ferrara 2005]{2005SSRv..116..625C} Ciardi, B., \&
  Ferrara, A.\ 2005, Space Science Reviews, 116: 625-705

\bibitem[Corbin et al. 2005]{LUDAR}
Corbin, M. et al. 2005, exploratory proposal to NASA

\bibitem[]{} Falcke, H. 2006, {\sl IAU JD12: Long Wavelength
Astrophysics}, 12, 16

\bibitem[Fan et al. (2006)]{Fan06}
Fan. X., Carilli, C., Keating, B. 2006, ARAA, 44, 415

\bibitem[Furlanetto \& Loeb 2002]{FL02}
Furlanetto, S, \&  Loeb, A. 2002, ApJ, 579, 1-9

\bibitem[Furlanetto et al. 2004]{FZH04}
Furlanetto, S.,  Zaldarriaga, M. Hernquist, L.
2004, ApJ, 613, 16-22

\bibitem[Furlanetto et al. 2006]{furl06}
Furlanetto, S.,  Oh, S., Briggs, F. 2006, 
Phys.Rep. in press.

\bibitem[Gnedin 2004]{Gnedin04}
Gnedin, N. 2004, ApJ, 610, 9-13

\bibitem[Gnedin \& Shaver 2003]{GS03}
Gnedin, N. \& Shaver, P. 2004, ApJ, 608, 611-621

\bibitem[]{} Gorgolewski, S. 1965, Astronautica Acta, New Series 11,
126, 130

\bibitem[Haiman et al. 2004]{haiman04}
Haiman, Z., Quartaert, E., Bower, G. 2004, ApJ, 612, 698-705
 
\bibitem[]{} Hankins, T., Ekers, R., O'Sulliva, J. 2000, in {\sl
First international workshop RADHEP,} eds. 
D. Saltzberg \& P. Gorham, New York: AIP, p. 168

\bibitem[]{} Heidemann, J. 2000, Ad.Sp.Rev.-26, 2, 343

\bibitem[]{} Hopkins, P., Doeleman, S., Lonsdale, C. 2003, AAS, 203,
4005 

\bibitem[Jarvis \& Rawlings 2005]{JR05} Jarvis, M. \& Rawlings,
S. 2005, New AR, 48, 1173

\bibitem[]{} Kassim, N., et al. 2006, in {\sl From Clark Lake to the
Long Wavelength Array}, ASP Conference Series, Vol. 345,
Eds N. Kassim, M. Perez, M. Junor, and P. Henning, 392

\bibitem[]{} Keshet, U., Waxman, E.,
\& Loeb, A.\ 2004, ApJ, 617, 281; New Astronomy Review, 48, 1119

\bibitem[]{}Kuiper, T., Jones, D., Mahoney, M., Preston, R.  1990, in
{\sl Astrophysics from the moon}, New York: AIP, p. 522

\bibitem[]{} Lane, W.; Cohen, A.; Cotton, W.D.;
Condon, J.; Perley, R.A.; Lazio, J.; Kassim, N.;
Erickson, W. 2004, SPIE, 5489, 354

\bibitem[]{} Lazio, J. et al. 2004, ApJ, 612, 511

\bibitem[]{} Lazio, J. et al. 2006, {\sl IAU JD12: Long Wavelength
Astrophysics}, 12, 62

\bibitem[]{} Loeb, A. 2006, in {\sl SAAS-Fee Winter School: First
Light}, Springer: Berlin, in press

\bibitem[]{} Loeb, A., \& Zaldarriaga, M. 2004, Phys. Rev. Lett.,
92, 1301

\bibitem[Madau et al. 1997]{MMR97}
Madau, P., Meiksin, A., Rees, M. 1997,
ApJ, 475, 429-444

\bibitem[Mcquinn et al. 2006]{mcquinn} Mcquinn, M. et al.
2006, ApJ, in press

\bibitem[]{} Mellema, G., Iliev, I., Pen, U.-L., Shapiro, P. 2006,
MNRAS, 372, 679

\bibitem[Morales \& Hewitt 2004]{MH03} Morales, M. \& Hewitt, J. 2004,
ApJ, 615, 7

\bibitem[Morales 2005]{Morales04}
Morales, M. 2005, ApJ, 619, 678

\bibitem[]{} Perley, R. 1999, in Synthesis Imaging in Radio Astronomy
II, PASP:San Francisco, eds. Taylor, Carilli, Perley, 180, 275

\bibitem[Santos et al. 2004]{2004ApJ...606..683S} Santos, M.~R.,
  Ellis, R.~S., Kneib, J.-P., Richard, J., \& Kuijken, K.\ 2004, ApJ,
  606: 683

\bibitem[Sethi 2005]{sethi05}
Sethi, S. 2005, MNRAS, 363, 818

\bibitem[]{} Vyshlov, A. 1974, Spaec Research, 16, 945

\bibitem[]{} Waon, G. et al. 1997, ESA Sci(97) 2

\bibitem[Wyithe et al. 2005]{WLB05}
Wyithe, J.S., Loeb, A., Barnes, D. 2005, ApJ, 634, 715

\bibitem[]{} Wouthuysen, S. 1952, AJ, 57, 31-33

\bibitem[Zaldarriaga et al. 2004]{ZFH04}
Zaldarriaga, M., Furlanetto, S., Henquist, L. 2004, ApJ, 608, 622

\end{thebibliography}
\end{document}